# ISOTOPIC EFFECTS IN NUCLEAR REACTIONS AT RELATIVISTIC ENERGIES

C. Sfienti[(1)], M. De Napoli[(2)], S. Bianchin[(1)], A.S. Botvina[(1)(6)], J. Brzychczyk[(4)],
A. Le Fèvre[(1)], J. Lukasik[(1)(9)], P. Pawlowski[(4)], W. Trautmann[(1)], P. Adrich[(1)],
T. Aumann[(1)], C.O. Bacri[(3)], T. Barczyk[(4)], R. Bassini[(5)], C. Boiano[(5)], A. Boudard[(7)],
A. Chbihi[(8)], J. Cibor[(9)], B. Czech[(9)], J.-E. Ducret[(7)], H. Emling[(1)], J. Frankland[(8)],
M. Hellström[(1)], D. Henzlova[(1)], K. Kezzar[(1)], G. Immè[(2)], I. Iori[(5)], H. Johansson[(1)],
A. Lafriakh[(3)], E. Le Gentil[(7)], Y. Leifels[(1)], W.G. Lynch[(10)], J. Lühning[(1)], U. Lynen[(1)],
Z. Majka[(4)], M. Mocko[(10)], W.F.J. Müller[(1)], A. Mykulyak[(11)], H. Orth[(1)], A.N. Otte[(1)],
R. Palit[(1)], A. Pullia[(5)], G. Raciti[(2)], E. Rapisarda[(2)], H. Sann[(1)†], C. Schwarz[(1)], H. Simon[(1)],
K. Sümmerer[(1)], C. Volant[(7)], M. Wallace[(10)], H. Weick[(1)], J. Wiechula[(1)], A. Wieloch[(4)]
and B. Zwieglinski[(11)]

**The ALADiN2000 Collaboration**
(1) Gesellschaft für Schwerionenforschung, D-64291 Darmstadt, Germany
(2) Dipartimento di Fisica dell'Università and LNS-INFN, I-95126 Catania, Italy
(3) Institut de Physique Nuclèaire, IN2P3-CNRS et Universitè, F-91406 Orsay, France
(4) M. Smoluchowski Institute of Physics, Jagiellonian Univ., Pl-30059 Krakòw, Poland
(5) Istituto di Scienze Fisiche, Università degli Studi and INFN, I-20133 Milano, Italy
(6) Inst. Nucl. Res., Russian Accademy of Science, Ru-117312 Moscow, Russia
(7) DAPNIA/SPhN, CEA/Saclay, F-91191 Gif-sur-Yvette, France
(8) GANIL, CEA et IN2P3-CNRS, F-14076 Caen, France
(9) H. Niewodniczanski Institute of Nuclear Physics, Pl-31342 Krakòw, Poland
(10) Department of Physics and Astronomy and NSCL, MSU, East Lansing, MI 48824, USA
(11) A. Soltan Institute for Nuclear Studies, Pl-00681 Warsaw, Poland

**Abstract**
A systematic study of isotopic effects in the break-up of projectile spectators at relativistic energies has been performed at the GSI laboratory with the ALADiN spectrometer coupled to the LAND neutron detector. Besides a primary beam of $^{124}$Sn, also secondary beams of $^{124}$La and $^{107}$Sn produced at the FRS fragment separator have been used in order to extend the range of isotopic compositions.
The gross properties of projectile fragmentation are very similar for all the studied systems but specific isotopic effects have been observed in both neutron and charged particle production. The breakup temperatures obtained from the double ratios of isotopic yields have been extracted and compared with the limiting-temperature expectation.

## 1. Introduction

Challenging motivations for isotopic studies in nuclear multifragmentation are derived from the importance of the density dependence of the symmetry-energy term of the nuclear equation of state for astrophysical applications and for effects linked to the manifestation of the nuclear liquid-gas phase transition.
Müller and Serot, in their seminal paper [1], have demonstrated that the two-fluid nature of nuclear matter has very specific consequences for the phase behavior in the coexistence region. Different isotopic compositions are predicted for the coexisting liquid and gas phases, with the gas being more neutron rich than the liquid in asymmetric (N ≠ Z) matter. This difference stems from the decrease in the symmetry energy in nuclear matter as the density is decreased. The expected magnitude of this density dependence, however, is model dependent and very poorly constrained by existing data [2].

The calculations of Müller and Serot are restricted to infinite matter with no Coulomb force included. In addition, the isotopic composition, in the calculations, is typically varied within a range of proton fractions $\rho_p/\rho$ = 0.3 to 0.5 whose limits are not easily accessible in experiments with heavy nuclei.

Theoretical studies for finite systems also indicate that the sequential decay of excited reaction products has a tendency to reduce some of the expected effects [3].

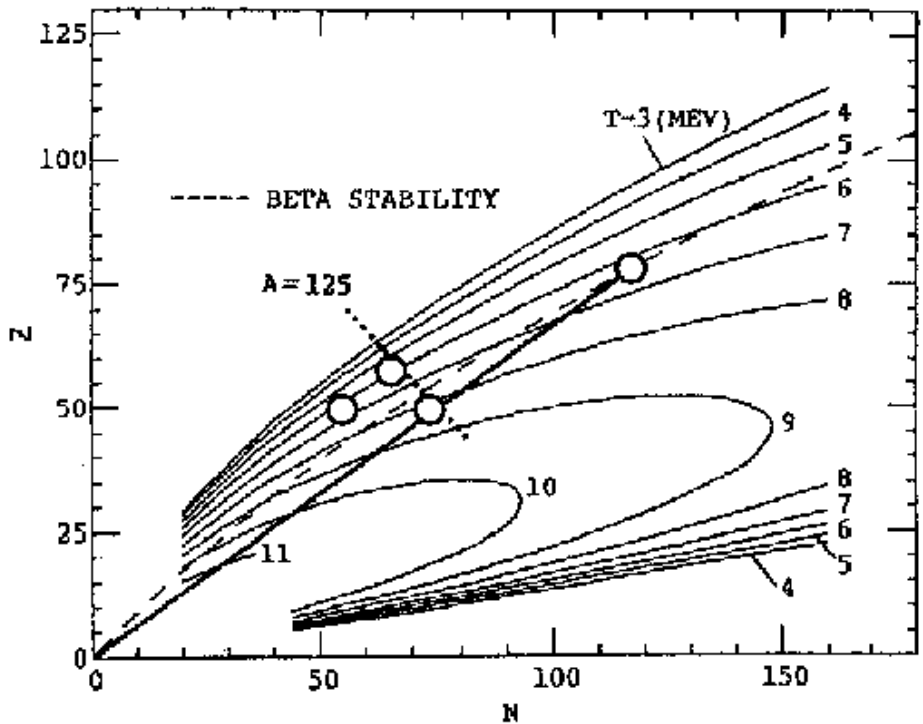

**Figure 1.** Location of the four studied projectiles in the plane of atomic number Z versus neutron number N. The contour lines represent the limiting temperatures according to [4], the dashed line gives the valley of stability, and the full line corresponds to the N/Z = 1.49 of $^{197}$Au.

If the phase transition for asymmetric nuclei still manifests itself as a plateau in the caloric curve [5] (i.e. the correlation between the temperature of the system and its excitation energy), this could be influenced by the degree of asymmetry as well as by the mass of the system undergoing the fragmentation.

It has been shown [4] that, due to the Coulomb pressure, there exists a limiting temperature which represents the maximum temperature at which nuclei are found to exist as self-bound objects in Hartree-Fock calculations. The dependence of the breakup temperature on the excitation energy could therefore be governed by the limiting temperature. Figure 1 shows the calculated limiting temperature as a function of the neutron and proton number [4]. As it is possible to see from the picture, in case of proton-rich nuclei, the phenomenon of vanishing limiting temperature is predicted, the limiting temperature decreasing with increasing proton fraction.

Clearly, new experiments exploring such phenomena are mandatory for having a better knowledge of the thermodynamics of a finite nucleus and its decay.

Recently a systematic investigation of projectile-spectator fragmentation has been undertaken at the ALADiN spectrometer at the GSI [6,7]: four different projectiles, $^{124}$Sn, $^{197}$Au, $^{124}$La and $^{107}$Sn, all with an incident energy of 600 AMeV on $^{116}$Sn and $^{197}$Au targets, have been studied. The two latter beams have been delivered by the FRagment Separator (FRS) of the GSI as products of the fragmentation of a primary $^{142}$Nd beam at 890 AMeV on a $^9$Be production target [7].

The necessity of low beam intensities for the best operational condition of the ALADiN setup ($\cong$2000 particles/sec), and the possibility of using a thick target in order to achieve high interaction rates are indeed conditions compatible with radioactive-ion-beam experiments. Moreover, the inverse kinematics offers the possibility of a threshold-free detection of all heavy fragments and residues and thus gives a unique access to the breakup dynamics.

The measurement of the charge and the momentum vector of all projectile fragments with Z$\geq$2 has been performed, with high efficiency and high resolution, with the TP-MUSIC IV detector [8]. Using the reconstructed values for the rigidity and pathlength, the charge of the particle measured by the TP-MUSIC IV, and the time-of-flight given by the TOF-Wall, the velocity and the momentum vector can be calculated for each detected charged particle. The knowledge of velocity and momentum allows then the calculation of the particle's mass.

Neutrons emitted in directions close to $\theta_{lab} = 0^o$, are detected with the Large-Area Neutron Detector (LAND) which covers about half of the solid angle required for neutrons from the spectator decay.

## 2. Gross Properties of multifragment decay

The gross properties of projectile fragmentation are very similar for all the studied systems.

The fragments emerging from the decay of the projectile spectators are well localised in rapidity. The distributions are peaked around a rapidity value very close to be beam rapidity and become narrower with increasing mass of the fragment.

The observed independence of the *Rise and Fall*, i.e. of the correlation between the multplicity of intermediate mass fragments with $Z_{bound}$, also for the unstable systems [6], confirms the hypothesis of equilibrium at freeze-out.

The shape of the charge distributions are as well similar for all the systems. For larger values of $Z_{bound}$, the charge distribution exhibits a U-shape. The heavy fragments are the residues of the lowly-excited projectile-spectators after the evaporation of light fragments and nucleons. In semi-central reactions, i.e. smaller values of $Z_{bound}$, the distribution broadens and flattens over nearly the full charge distribution. For still lower values of $Z_{bound}$, the charge distribution becomes steep. This is consistent with the system disassembling into predominantly lighter fragments.

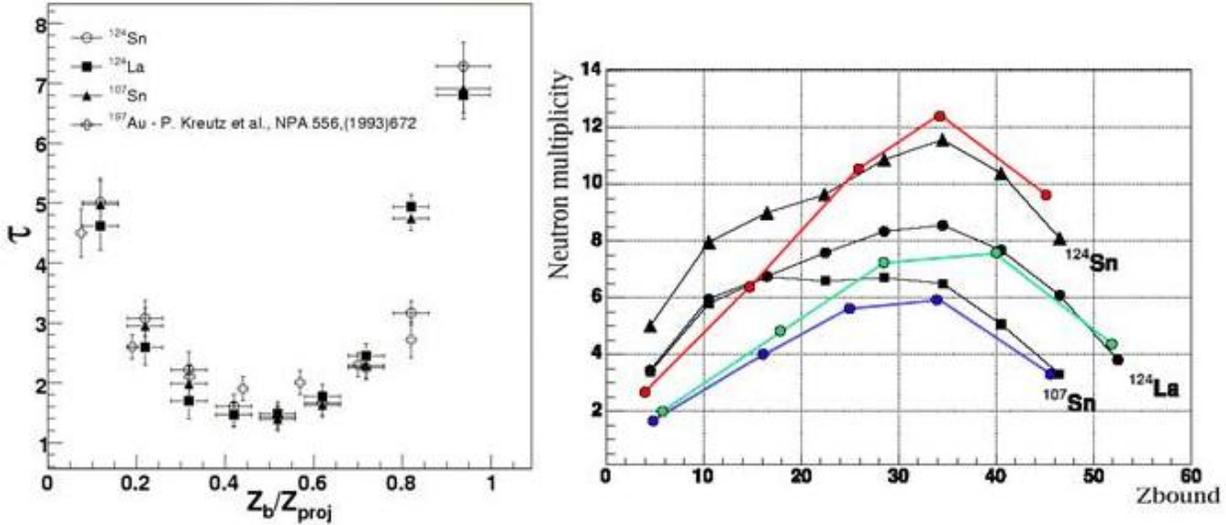

**Figure 2**. (Left Panel) The extracted $\tau$ parameters as a function of normalized $Z_{bound}$ for $^{124}$La, $^{124}$Sn and $^{107}$Sn at 600 AMeV, compared with earlier data for $^{197}$Au obtained at the same energy. (Right Panel) Neutron multiplicities as obtained from the LAND detector for all the studied systems (colored symbols are SMM predictions).

The charge distributions have been fitted with a power-law parameterization, $\sigma(Z) \propto Z^{-\tau}$, in the charge range $3 \leq Z \leq 15$. The power-law parameters $\tau$ (Figure 2 – left panel) allow to follow the transition from the U-shape to a pure exponential spectrum and reach the minimum value in the $Z_{bound}$ range which corresponds to the maximum production of IMF, i.e. in the multifragmentation region. They follow a nearly universal curve almost independent of the isotopic composition of the original spectator system.

Specific isotopic effects, even though small, can nevertheless be observed: in particular, the hierarchy of $\tau$ for the neutron-poor $^{124}$La and $^{107}$Sn and neutron-rich $^{124}$Sn and $^{197}$Au systems for $Z_{bound}/Z_{proj}>0.5$ is opposite to the standard predictions of the Statistical Multifragmentation Model SMM [9]. It can, however, be explained with a weak isotopic dependence of the surface-term coefficient in the liquid-drop description of the fragment masses at low excitation energy which gradually disappears with increasing excitation of the fragmenting system [10].

Specific isotopic effects are as well found in the reconstructed neutron multiplicities measured with the LAND detector [11] (Figure 2 - right panel). For peripheral collisions the values of the multiplicity depend on the number of available neutrons in the entrance channel. Going towards central collisions the number of emitted neutrons is progressively determined by the N/Z in the entrance channel. In a preliminary comparison with SMM calculation (colored points in Figure 2) a promising agreement has been observed.

Neutrons will be important for establishing the mass and energy balance and in particular for calorimetry. In this respect it is crucial to identify the spectator neutrons and to distinguish them from the fireball ones. From a preliminary analysis of their rapidity distributions, the corresponding spectator sources have been identified. They are characterized by temperatures up to 4 MeV, possibly caused by large contributions from evaporation.

## 2. Structure and Memory Effects in particle production

The mass resolution obtained for projectile fragments entering into the acceptance of the ALADiN spectrometer is about 3% for fragments with Z=3 and decreases to 1.5% for Z$\geq$6 [7]. Masses are thus individually resolved for fragments with atomic number Z$\leq$10. The elements are resolved over the full range of atomic numbers up to the projectile Z with a resolution of $\Delta Z \leq 0.2$ obtained with the TP-MUSIC IV detector. The mean N/Z of the mass distributions of light fragments in the range $3 \leq Z \leq 13$ for two different $Z_{bound}$ cuts is presented in Figure 3.

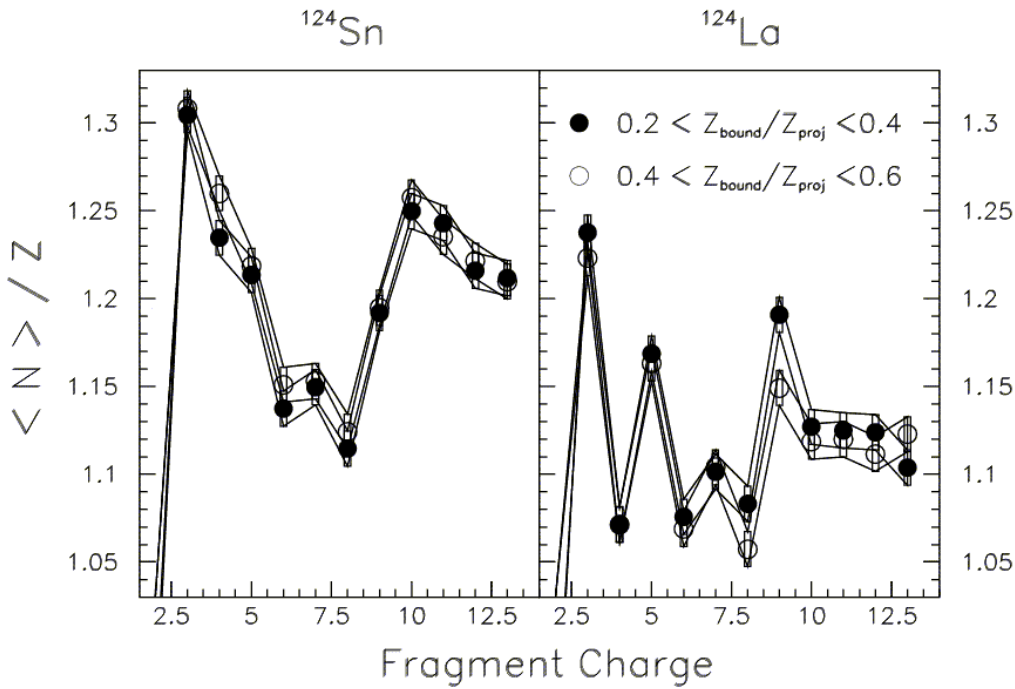

**Figure 3.** Mean values <N>/Z of light fragments with $3 \leq Z \leq 13$ produced in the fragmentation of $^{124}$Sn and $^{124}$La at 600 A MeV for two different bins in $Z_{bound.}$

The values obtained for $^{124}$Sn are larger than those for $^{124}$La or $^{107}$Sn (not shown) as expected from the different N/Z of the original projectiles. Their odd-even variation is, however, much more strongly pronounced for the neutron-poor cases. The strongly bound α-type nuclei (even-even N=Z) attract a large fraction of the product yields during the secondary evaporation stage. This effect is, apparently, larger if already the hot fragments are close to N=Z symmetry, as it is expected for the fragmentation of $^{124}$La and $^{107}$Sn [12]. Inclusive data obtained with the FRS fragment separator at GSI for $^{238}$U [13] and $^{56}$Fe [14] fragmentations on titanium targets at 1 A GeV bombarding energy confirm that the observed patterns are very systematic, exhibiting at the same time nuclear structure effects characteristic for the isotopes produced and significant memory effects of the isotopic composition of the excited system by which they are emitted. This has the consequence that, because of its strong variation with Z, the neutron-to-proton ratio <N>/Z is not a useful observable for studying nuclear matter properties. For this purpose, techniques, such as the isoscaling [7], will have to be used which cause the nuclear structure effects to cancel out. A precise modeling of these secondary processes is, therefore, necessary for quantitative analyses.

Another interesting feature of the mass distributions predicted by SMM is their dependence, in the case of the proton-rich systems, on the excitation energy of the system [10]. This difference arises from the dependence of the number of neutrons in light fragments on the Z spectrum. For the neutron-rich $^{124}$Sn system, on the other hand, the distributions should be independent of the excitation energy. From a first qualitative comparison with the model prediction of the mass distributions for different $Z_{bound}$ cuts, however, we have not observed any noticeable variation in the mean N/Zs for the $^{124}$La system (Figure 3). Also in this case it will be crucial to precisely model the sequential decay in order to clarify whether it has washed out the expected effect.

## 3. Limiting temperature

As previously mentioned, for proton-rich systems a rapid drop of the limiting temperature has been theoretically predicted (Figure 1) because of the increasing Coulomb pressure [4]. On the other hand, SMM calculations predict nearly isospin-invariant temperatures for the coexistence region [10]. Therefore, in order to distinguish whether the breakup temperature is determined by the binding properties of the excited hot nuclear system or by the phase space accessible to it by fragmentation we have analyzed the temperature for the studied neutron-poor and neutron-rich systems.

The temperatures used in the caloric curve studies [5,15] were deduced from a double ratio of isotopic yields. Under the assumptions of low density and chemical equilibrium a grand canonical treatment [16] yields for the double ratio:

$$\frac{Y_{(A_1,Z_1)}/Y_{(A_2,Z_2)}}{Y_{(A_3,Z_3)}/Y_{(A_4,Z_4)}} = a \exp\left(\frac{\Delta B}{T}\right)$$

with ΔB being the double difference of the binding energies of the chosen isotopes and $a$ a statistical factor containing spin and mass terms. Furthermore, for best sensitivity the binding energy difference ΔB should be comparable or larger than the temperatures to be measured [17].

The thermometry with the $^{3,4}$He isotope pairs used in the nuclear caloric curve [5] benefits from the large difference ΔB = 20.6 MeV of the binding energies of these two nuclei but it is not the only choice. There are also other combinations of isotopes which can be expected to provide the necessary sensitivity for the measurement of temperatures in the MeV range. In a recent systematic study of different isotopic thermometers for spectator fragmentation [18] $T_{BeLi}$, $T_{CLi}$, and $T_{CC}$ have been analyzed in detail. In particular, it has been found that the rise at small $Z_{bound}$, i.e. high excitation energy, previously observed with $T_{HeLi}$ is well reproduced by most thermometers, including $T_{BeLi}$ which is derived from the $^{7,9}$Be and $^{6,8}$Li isotope ratios. This is an indication that the rise is not necessarily related to a particular behavior of either $^3$He or $^4$He. An overall good agreement between the different temperature observables has been obtained with the exception of those containing carbon isotopes. In the latter cases, the apparent temperature values remain approximately constant with values between 4 and 5 MeV.

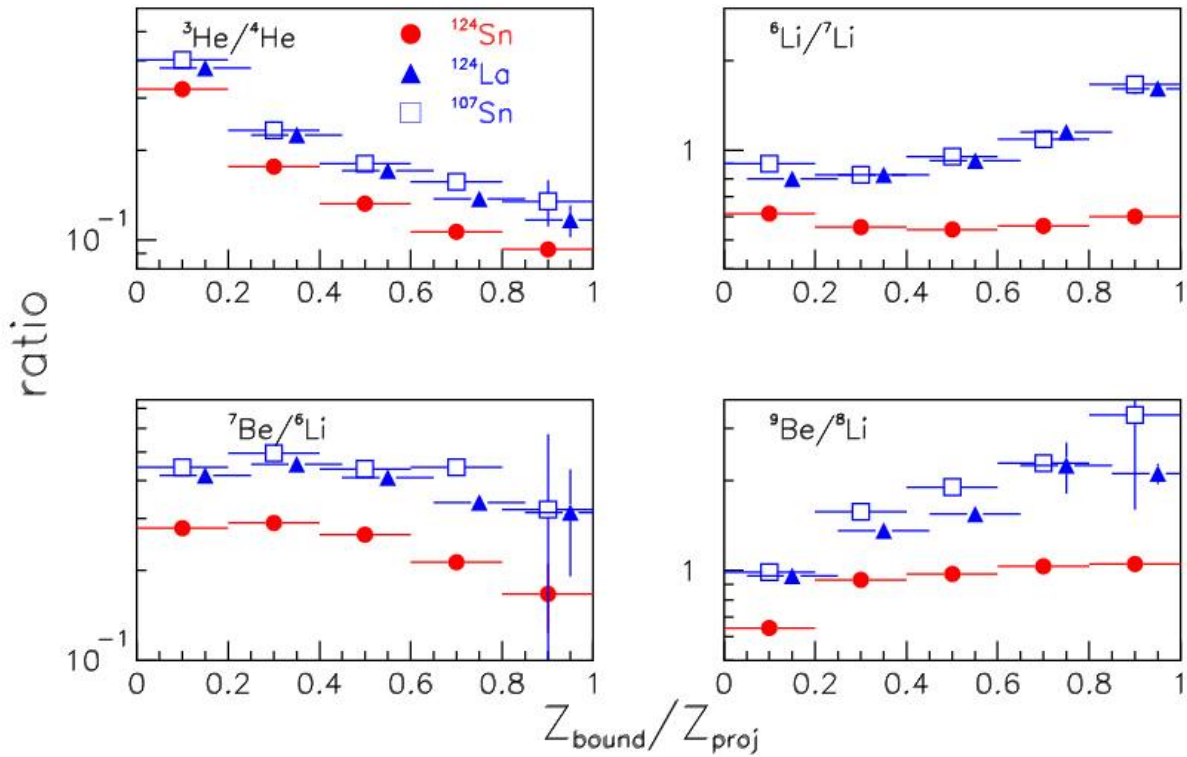

**Figure 4.** Measured isotopic yield ratios as a function of $Z_{bound}/Z_{proj}$ for the neutron-rich $^{124}$Sn and the neutron-poor $^{124}$La and $^{107}$Sn systems.

Measured isotope ratios as a function of $Z_{bound}$ for the $^{124}$Sn, $^{124}$La and $^{107}$Sn systems are shown in Figure 4. A very interesting dependence on the isotopic composition of the spectator is observed for all analyzed ratios with the exception of the $^3$He/$^4$He ratio, exhibiting a reduced difference between the neutron-rich and neutron-poor systems. The strong sequential decay into α particles could explain the difference between these two behaviors. Therefore, also in this case, the structure effects responsible for the observed odd-even variation in the N/Z/s of intermediate-mass fragments (Figure 3) play the major role.

For the $^{124}$Sn projectile spectator, moreover, the ratios, with the exception of $^3$He/$^4$He, are almost independent of $Z_{bound}$. For the mentioned ratio the decrease between the most central and the most peripheral collisions is about a factor 4 whereas for example, for the $^9$Be/$^8$Li ratio the total variation amounts hardly to a factor of 2. There seems to be a clear correlation between the difference of the binding energies of the involved isotopes and $Z_{bound}$, i.e. excitation energy of the system [18]: the higher the first the stronger the variation of the corresponding ratio with $Z_{bound}$.

On the other hand, in the case of the proton-rich systems all ratios exhibit a strong dependence on the excitation energy deposited in the system.

From the measured ratios the isotopic temperatures have been extracted and an overall agreement with the previous systematics [18] have been obtained for the $T_{HeLi}$, $T_{BeLi}$, $T_{CLi}$, and $T_{CC}$ thermometers.

In Figure 5 in particular, the apparent $T_{HeLi}$ and $T_{BeLi}$ temperatures for the $^{124}$Sn and $^{124}$La,

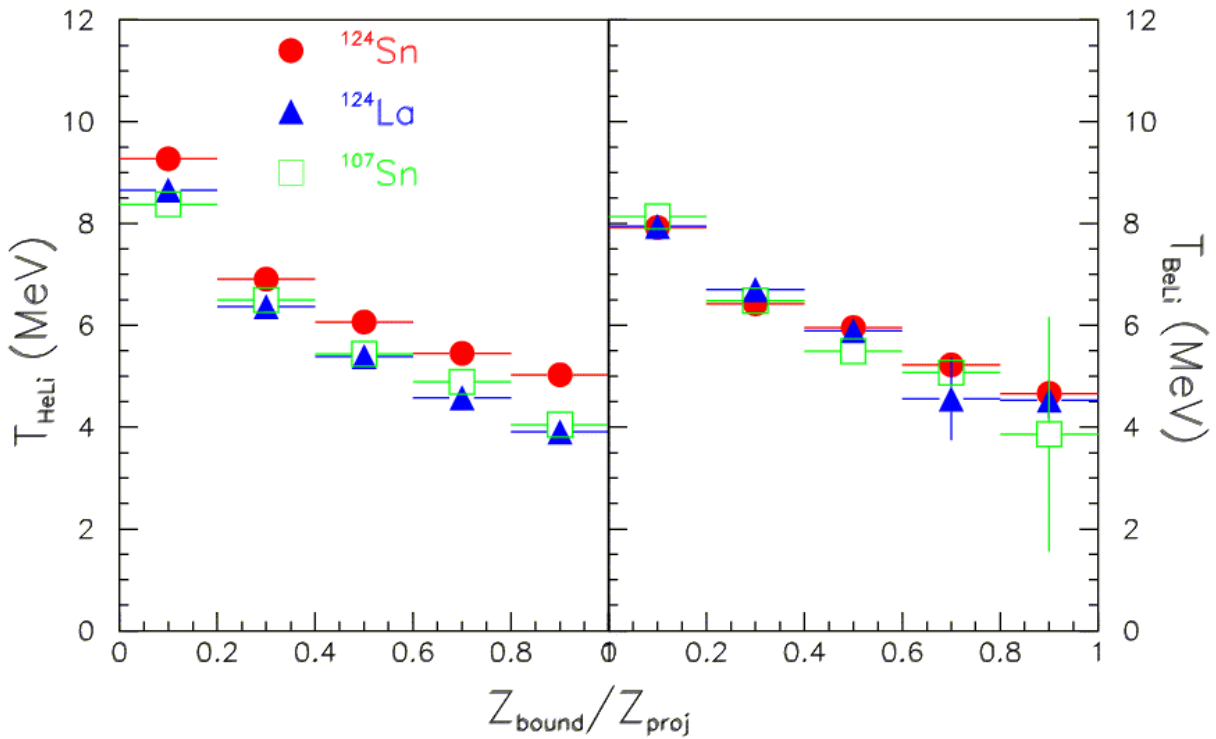

**Figure 5.** Isotopic temperatures $T_{HeLi}$ and $T_{BeLi}$ as a function of $Z_{bound}/Z_{proj}$ for $^{124}$Sn and $^{124}$La, $^{107}$Sn projectile spectators.

$^{107}$Sn spectator systems are reported. By comparing the two systems with the same mass but different isospin-content, the average difference of the obtained $T_{HeLi}$ temperatures is 0.7±0.1 MeV whereas almost no difference (hardly 0.1 MeV in average) between the two systems is observed in the case of the $T_{BeLi}$ thermometer. The small difference observed in the case of $T_{HeLi}$ is caused by the fact that the dependence of the $^6$Li/$^7$Li ratio (Figure 4) on the isotopic composition of the system is not compensated by the weak dependence of the $^3$He/$^4$He ratio. In the case of $T_{BeLi}$ both isotopic ratios $^{7,9}$Be and $^{6,8}$Li exhibit a dependence on the N/Z of the original projectiles, which cancels out in the double ratio. The invariance of the isotopic temperature with the isotopic composition of the system is inconsistent with the limiting temperature predictions [4] of 2 MeV for the $^{124}$Sn and $^{124}$La systems (Figure 1) and seems to favor a statistical interpretation [10].

## 4. Conclusions

Isotopic effects in the break-up of projectile spectators at relativistic energies have been reported. The gross properties of projectile fragmentation are very similar for all the studied systems. Specific isotopic effects, even though small, can nevertheless be observed: in particular, the inversion in the hierarchy of the τ exponential parameter of the charge distribution can be explained with a weak isotopic dependence of the surface-term coefficient of the nuclear equation of state [10].

The mean N/Zs of the isotope distributions of light fragments exhibit as well specific isotopic effects. In particular, the observed odd-even variation is much more strongly pronounced for the neutron-poor cases and could be explained in terms of a simultaneous concurrence of both structure effects characteristic for the isotopes produced and memory effects of the isotopic composition of the excited system from which they are emitted.

From the double ratios of Z≤4 isotopes, the isotopic temperatures have been determined. The small dependence (of about 0.7 MeV) observed in the $T_{HeLi}$ thermometer could be due to the influence of structure effects in the sequential decay. The invariance with the isotopic composition in the entrance channel is inconsistent with the limiting-temperature predictions and seems to favors the statistical interpretation. On the other hand, the limiting-temperature concept reproduces nicely the mass dependence of the caloric curves [15] and only seems to fail when applied to the isospin degree of freedom.

It should be noted that most of the limiting-temperature calculations are made for beta stable nuclei [19] while it was not explicitly tested whether the studied systems are still near the stability at breakup.

The weak N/Z dependence of the breakup temperatures measured in this experiment shows that this is not a major point of concern. The same observation, on the other hand, is not compatible with the strong Coulomb effect predicted by Besprosvany and Levit [4]. This open point in the connection between limiting temperatures for heavy nuclei and breakup temperatures of fragmenting nuclear systems will require an improved understanding.

*C.Sf. acknowledges the receipt of an Alexander-von-Humboldt fellowship. This work was supported by the European Community under contract No. RII3-CT-2004-506078 and HPRI-CT-1999-00001 and by the Polish Scientific Research Committee under contract No. 2P03B11023.*